\documentclass[8.5pt,twoside,twocolumn]{article}
\usepackage[super,sort&compress,comma]{natbib} 
\usepackage{times,mathptmx}

\usepackage{sectsty}
\usepackage{balance} 

\usepackage{graphicx} 

\usepackage{lastpage}

\usepackage[format=plain,justification=raggedright,singlelinecheck=false,font=small,labelfont=bf,labelsep=space]{caption} 
\usepackage{fancyhdr}
\usepackage{color}
\usepackage{amssymb}
\usepackage{amsmath}

\begin{document}

\thispagestyle{plain}
\fancypagestyle{plain}{
\fancyhead[L]{\includegraphics[height=8pt]{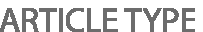}}
\fancyhead[C]{\hspace{-1cm}\includegraphics[height=20pt]{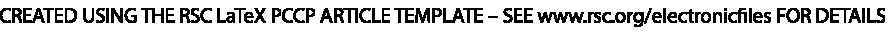}}
\fancyhead[R]{\includegraphics[height=10pt]{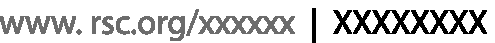}\vspace{-0.2cm}}
\renewcommand{\headrulewidth}{1pt}}
\renewcommand{\thefootnote}{\fnsymbol{footnote}}
\renewcommand\footnoterule{\vspace*{1pt}%
\hrule width 3.4in height 0.4pt \vspace*{5pt}} 
\setcounter{secnumdepth}{5}

\makeatletter 
\def\subsubsection{\@startsection{subsubsection}{3}{10pt}{-1.25ex plus -1ex minus -.1ex}{0ex plus 0ex}{\normalsize\bf}} 
\def\paragraph{\@startsection{paragraph}{4}{10pt}{-1.25ex plus -1ex minus -.1ex}{0ex plus 0ex}{\normalsize\textit}} 
\renewcommand\@biblabel[1]{#1}
\renewcommand\@makefntext[1]%
{\noindent\makebox[0pt][r]{\@thefnmark\,}#1}
\makeatother 
\renewcommand{\figurename}{\small{Fig.}~}
\sectionfont{\large}
\subsectionfont{\normalsize} 

\fancyfoot{}
\fancyfoot[LO,RE]{\vspace{-7pt}\includegraphics[height=9pt]{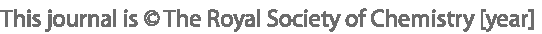}}
\fancyfoot[CO]{\vspace{-7.2pt}\hspace{12.2cm}\includegraphics{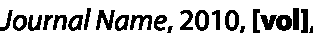}}
\fancyfoot[CE]{\vspace{-7.5pt}\hspace{-13.5cm}\includegraphics{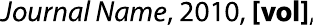}}
\fancyfoot[RO]{\footnotesize{\sffamily{1--\pageref{LastPage} ~\textbar  \hspace{2pt}\thepage}}}
\fancyfoot[LE]{\footnotesize{\sffamily{\thepage~\textbar\hspace{3.45cm} 1--\pageref{LastPage}}}}
\fancyhead{}
\renewcommand{\headrulewidth}{1pt} 
\renewcommand{\footrulewidth}{1pt}
\setlength{\arrayrulewidth}{1pt}
\setlength{\columnsep}{6.5mm}
\setlength\bibsep{1pt}

\twocolumn[
  \begin{@twocolumnfalse}
\noindent\LARGE{\textbf{Soft electrowetting}}

\vspace{0.6cm}

\noindent\large{\textbf{R. Dey\textit{$^{\ast}$ $^{a, b\dagger}$} , M. van Gorcum,\textit{$^{c\dagger}$} F. Mugele,\textit{$^{b}$} and J. H. Snoeijer\textit{$^{c}$}}}\vspace{0.5cm}

\noindent\textit{\small{\textbf{Received Xth XXXXXXXXXX 20XX, Accepted Xth XXXXXXXXX 20XX\newline
First published on the web Xth XXXXXXXXXX 20XX}}}



\noindent \normalsize{
Electrowetting is a commonly used tool to manipulate sessile drops on hydrophobic surfaces. By applying an external voltage over a liquid and a dielectric-coated surface, one achieves a reduction of the macroscopic contact angles for increasing voltage. The electrostatic forces all play out near the contact line, on a scale of the order of the thickness of the solid dielectric layer. Here we explore the case where the dielectric is a soft elastic layer, which deforms elastically under the effect of electrostatic and capillary forces. The wetting behaviour is quantified by measurements of the static and dynamic contact angles, complemented by confocal microscopy to reveal the elastic deformations.  Even though the mechanics near the contact line is highly intricate, the macroscopic contact angles can be understood from global conservation laws in the spirit of Young-Lippmann. The key finding is that, while elasticity has no effect on the static electrowetting angle, the substrate's viscoelasticity completely dictates the spreading dynamics of electrowetting. }
\vspace{0.5cm}
 \end{@twocolumnfalse}
  ]

\footnotetext{$\dagger$~Both authors contributed equally}
\footnotetext{\textit{$^{a}$~Max Planck Institute for Dynamics and Self-Organization, Am Fassberg 17, 37077 Goettingen, Germany; E-mail: ranabir.dey@ds.mpg.de}}
\footnotetext{\textit{$^{b}$~Physics of Complex Fluids Group, Faculty of Science and Technology, University of Twente, P.O. Box 217, 7500AE Enschede, The Netherlands.}}
\footnotetext{\textit{$^{c}$~Physics of Fluids Group, Faculty of Science and Technology, University of Twente, P.O. Box 217, 7500AE Enschede, The Netherlands E-mail: j.h.snoeijer@utwente.nl}}

\section{Introduction}

\begin{figure*}
\centering
\includegraphics[width=120mm]{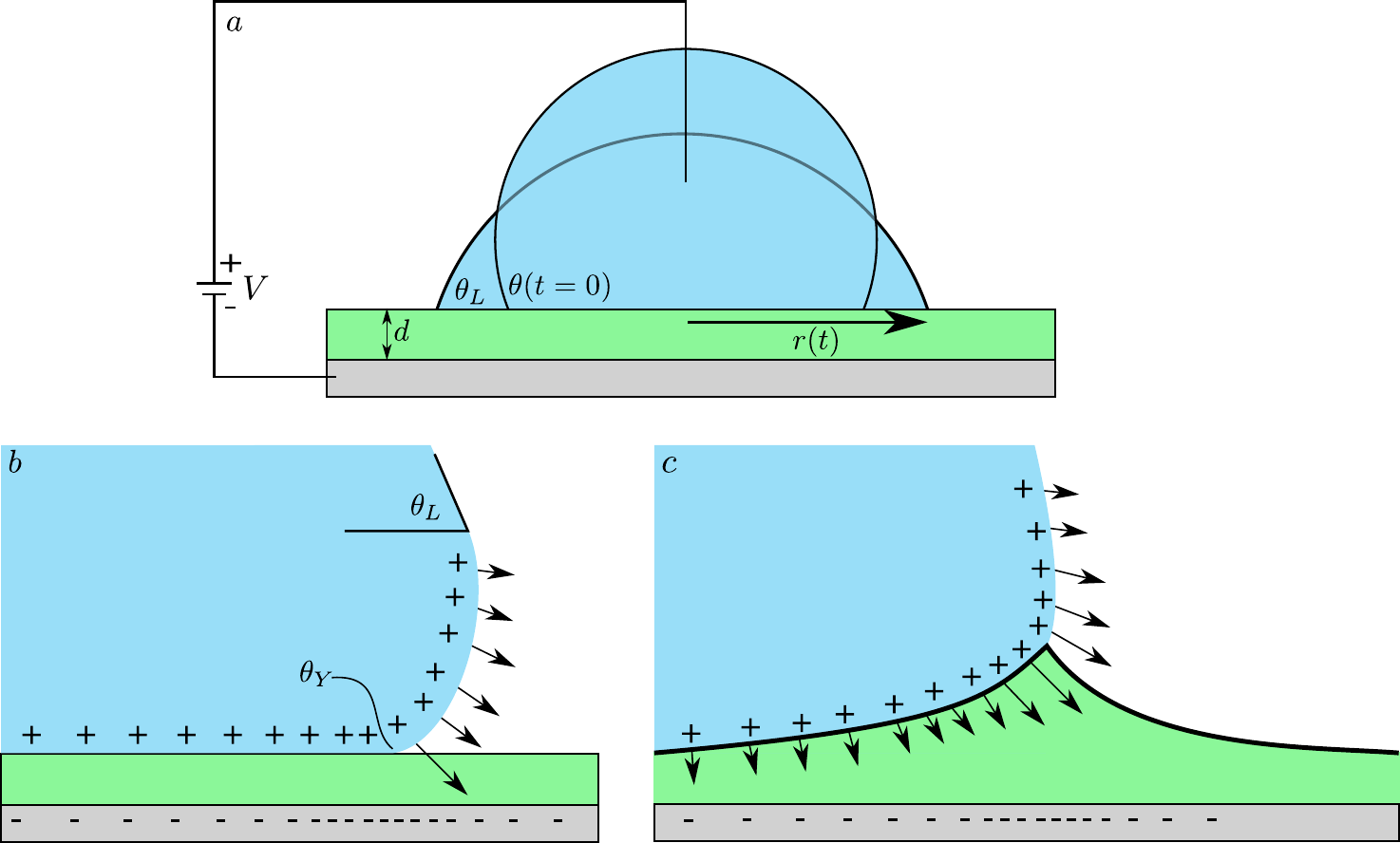}
\caption{Schematic view of electrowetting. (a) Macroscopic view: An electric potential is applied between the drop and the surface causing the macroscopic angle to decrease to the Lippmann angle $\theta_L$. (b) Microscopic view: At the contact line one recovers Young's angle $\theta_Y$. Electrostatic effects appear as a Maxwell stress on the surface charges along the liquid-vapor and liquid-solid interfaces. The balance of Maxwell stress and capillary pressure at the liquid-vapour interface induces a curvature of the liquid, deforming the interface from the microscopic $\theta_Y$ to the macroscopic Lippmann angle $\theta_L$. (c) Sketch of electrowetting on a soft dielectric. At the contact line, one expects a Neumann balance of the three surface tensions. The substrate shape follows from Maxwell stress, capillary pressure due to solid surface tension, and its bulk elasticity.}
\label{fgr:schematic}
\end{figure*}

The application of an external electrical voltage to a conducting sessile droplet, resting on a hydrophobic dielectric film covering an electrode, results in enhanced wetting of the droplet on the dielectric film (figure 1 a). Such electrically controlled partial wetting of a sessile liquid drop on a dielectric film is referred to as `electrowetting-on-dielectric' (EWOD) or simply `electrowetting' \cite{Mugele2005,mugele2018electrowetting}. Drop manipulation using electrowetting has a wide range of applications inlcuding lab-on-a-chip devices \cite{Fair2007}, optofluidic displays \cite{hayes2003video, heikenfeld2009electrofluidic}, optofluidic lenses \cite{mishra2016recent}, energy harvesting systems\cite{krupenkin2011reverse} and bio-analytical sample preparation \cite{miller2009digital,kudina2016maldi}. Moreover, it has been used as a general tool to investigate fundamental aspects of wetting of complex surfaces (see e.g. \cite{ shamai2008water, Ruiter2015, Lomax2016,hao2014electrowetting}) and viscoelastic fluids \cite{banpurkar2008electrowetting}.

The change in the macroscopic contact angle of the sessile drop under the applied electrical voltage can be understood by means of an energy minimization approach \cite{bruno1993electrocapillarite, Mugele2005,mugele2018electrowetting}. At zero voltage, the free energy of the system is given by the surface energies $\gamma$, $\gamma_{SL}$, $\gamma_{SV}$, respectively referring to the liquid-vapour, solid-liquid, and solid-vapour interfacial energies. The application of a voltage adds an electrostatic component to the free energy. This electrostatic contribution to the free energy originates from the difference between the energy stored in the dielectric and that stored in the external charging source (or the battery), as shown in figure \ref{fgr:schematic} a. Assuming that the droplet is a perfect conductor and has a size much larger than the dielectric thickness $d$, the net electrostatic free energy per unit wetted area reads $-\frac{\epsilon\epsilon_0 V^2}{2d}$, which effectively lowers the solid-liquid interfacial energy. Here $\epsilon$ is the relative permittivity of the dielectric, $\epsilon_0$ is the permittivity of free space, and $V$ is the applied voltage. After incorporating this electrostatic contribution in the total free energy of the system, and minimizing, we obtain the Young-Lippmann equation \cite{bruno1993electrocapillarite, Mugele2005,mugele2018electrowetting}:

\begin{equation}
\label{eq:electrowetting}
\cos \theta_{L} =\frac{\gamma_{SV}-  \left( \gamma_{SL}-\frac{\epsilon\epsilon_0 V^2}{2d}\right)}{\gamma}=\cos \theta_Y+\frac{\epsilon\epsilon_0}{2\gamma d}V^2.
\end{equation}
Equation (\ref{eq:electrowetting}) can be viewed as a modified form of Young's equation, which includes the electrostatic reduction of the effective wetted surface free energy. At vanishing voltage Young's angle $\theta_Y$ is recovered, while at finite $V$ one observes the Lippmann angle $\theta_{L}$. The dimensionless combination $\eta = \frac{\epsilon\epsilon_0 V^2}{2\gamma d}$  is often referred to as the electrowetting number \cite{Mugele2005,mugele2018electrowetting}. 

The Young-Lippmann equation can be also derived from a force balance on a liquid control volume very close to the contact line \cite{jones2002relationship}. On rigid surfaces, it turns out that the true microscopic angle is again given by Young's law~\cite{Buehrle2003}, i.e. we observe $\theta_Y$, as is sketched in figure \ref{fgr:schematic} b. However, the presence of surface charges gives rise to a Maxwell stress, pulling on the liquid-vapor interface. This Maxwell stress must be balanced by the Laplace pressure, which leads to a region of large curvature of the droplet surface near the contact line.  The Maxwell stress is distributed over a region having a length scale comparable to the dielectric thickness $d$. Hence, when the droplet length scale is much larger than $d$, the increased curvature (bending) of the droplet surface close to the contact line manifests in a lowering of the macroscopic contact angle $\theta_L$ as sketched in figure \ref{fgr:schematic} b. It is the variation of this macroscopic $\theta_L$ with $V$ that is given by Equation (\ref{eq:electrowetting}). 

In this paper we investigate how the aforementioned electrowetting scenario is altered if the dielectric layer is made soft. A sketch of `soft electrowetting' is shown in figure \ref{fgr:schematic} c. Wetting on soft surfaces leads to elastic deformations, altering the geometry near the contact line by creating a ridge on the soft substrate \cite{MDSA12b,Style2012a,Limat2012a,Lub14,BostwickSM14,style2017elastocapillarity}. The size of this wetting ridge can be of the order of the elastocapillary length $\gamma/G$, where $G$ is the shear modulus of the dielectric layer. The geometry of the wetting ridge is dictated by a vectorial balance of surface tensions, known as Neumann's law \cite{neumann1894vorlesungen,MDSA12b,Style2012a,Limat2012a}. However, in cases where the drop size is much larger than $\gamma/G$, the liquid contact angle typically remains at $\theta_Y$ \cite{Style2012a,Lub14}. Interestingly, for soft PDMS layers, the elastocapillary length can easily be tens of microns. For the physical situations when $\gamma/G$ is comparable to $d$, or larger, one thus expects an interplay between elasticity and electrostatics. This is of particular interest, since the highly localised forcing of electrowetting provides a way to actively probe the mechanics of the elastic wetting ridge.

The combination of the elasticity and electrowetting has been studied recently \cite{Dey2017, Dey2015}. These works showed experimentally that at a particular value of the applied voltage the macroscopic contact angle increases with increasing softness of the dielectric film \cite{Dey2017, Dey2015}, suggesting the possibility of a departure from the classical Young-Lippmann equation. This was complemented by an approximate theoretical description for the static shape of the wetting ridge as a function of voltage \cite{Dey2017}, which is still to be verified experimentally. However, the dynamics during electrowetting, and its effect on the shape of the wetting ridge remain unexplored. In this manuscript we therefore experimentally investigate the statics and dynamics of soft electrowetting. We perform a systematic study of the macroscopic contact angles of the liquid, as a function of voltage $V$ and contact line speed $U$. This is complemented with confocal microscopy results, where we show the shape of wetting ridges in the presence of strong Maxwell stresses.

\section{Methods}
Electrowetting on soft solids is explored using a sessile drop on an ITO coated glass on which we spincoat a thin layer of CY52-276 A/B PDMS gel. The gel has a shear modulus $G \sim1$ kPa and exhibits a viscoelastic rheology that is described in detail in \cite{KarpNcom15}. In particular, the loss modulus exhibits a power-law frequency dependence $G''/G = (\omega \tau)^n$, where $\tau \sim 0.1$ and the exponent $n \sim 0.55$.~\cite{KarpNcom15} The spincoating is done by first pouring the uncured gel on the substrate, spreading it out to the edges of the glass, and then subsequently spincoating for 40s at 30rps. The samples are then cured by heating these at $70^\circ$C for 30 minutes. This results in a dielectric gel layer thickness of $d=22\pm1\mu$m, measured using reflectometry with an Ocean Optics HR2000+ spectrometer and a HL-2000-FHSA halogen light source.
For the subsequent electrical connections, the ITO surface below the gel coating is connected to a wire by scratching off the gel in a small area and using silver paste to glue the wire to the surface. Norland Optical Adhesive is applied on top of the connection, and subsequently cured using UV light to fixate the wire. For the experiments, we use 1mM KCl solution in deionised water as a conductive liquid with a surface tension of $72$ mN/m. This gives an elastocapillary length $\gamma/G \sim 70 \, \mu$m, which is larger than the dielectric thickness $d$. Hence, we expect a large deformation of the soft substrate that is at least similar in magnitude to the dielectric layer thickness. 

To measure the dielectric constant of the soft film we perform an independent calibration of $\epsilon$. For this, a glass slide is coated with platinum in the shape of a square of known area. This glass slide is then placed below another platinum coated slide, with $0.55\pm0.01$ mm glass spacers in between. The gap is filled with the PDMS gel, and the capacitance of the system is measured using an HP 4194A impedance gain phase analyzer. During the measurement the entire system is wrapped in aluminium foil as a Faraday cage to prevent EM interference. From the capacitive measurement, we determine the  dielectric constant as $\epsilon=\frac{Cd}{\epsilon_0A}$. Here $C$ is the measured capacitance of $23\pm1$ pF, $d=0.55\pm0.01$ mm is the thickness of the dielectric layer for these measurements, $\epsilon_0$ is the permittivity of vaccuum, and $A=415\pm4$ mm$^2$ is the capacitance area. This gives the value of the dielectric constant of the gel as $\epsilon=3.4\pm0.3$. 

Dataphysics OCA15 is used to measure the contact angle variations. We use a Hamilton gas tight $500 \; \mu$l syringe to deposit the drop on the soft dielectric film. The metal needle is kept inserted inside the drop, and is connected to a function generator and an amplifier in order to apply a DC voltage between the drop and the ITO surface (grounded) as shown schematically in figure \ref{fgr:schematic}. It must be noted here that the maximum applied voltage is limited by the breakdown of the dielectric film beyond a threshold voltage value ($V_{max} \sim 250$ V). On quickly inflating the drop ($\sim5\mu$l/s), the drop spreads while the contact angle $\theta$ relaxes towards its equilibrium value $\theta_L$ under the applied voltage. This is tracked as shown in figure \ref{fgr:schematic} a. Typically the relaxation phase lasts for $\sim10$ minutes. 

The dynamics is quantified by measuring the contact angle $\theta(t)$ and the corresponding position of the contact line $r(t)$ (figure 1 a). This is done using a MATLAB code that detects the edge of the drop for each frame. The edge above the baseline is then fitted with a polynomial fit, on both left and right sides of the drop. The intersection point of the polynomial fit with the baseline then gives the position of the contact line, and the angle between the baseline and the polynomial fit at the intersection point gives the contact angle. The instantaneous contact line speed $U$ is subsequently extracted using a linear regression on the measured position of the contact line. The equilibrium angle $\theta_{L}$ is found by extrapolating the $\theta$ vs. $U$ data for vanishing velocity, and is cross-checked by finding the contact angle when the contact line speed reaches 0 from the data recorded using the OCA. This yields the same result because the contact angle hysteresis of the PDMS gel is negligible \cite{xu2017direct,Snoeijer2018}.

In a second set of experiments we focus on the shape of the wetting ridge in close proximity of the contact line under electrowetting. For this we use confocal microscopy. For confocal microscopy we use two fluoroscent dyes-- DFSB-K175 (Risk Reactor) in the gel and Alexa Fluor$^ \textup{TM}$ 647 (ThermoFisher SCIENTIFIC) in the drop. The dye DFSB-K175 is mixed with the gel in the volume ratio of $50 \; \mu$l per ml. In order to index match the droplet liquid with the gel, glycerol is mixed with deionized water in a container containing fully cured gel at the bottom until the interface between the liquid and the gel optically disappeared. Alexa Fluor 647 is added in a ratio of $0.2$mg per ml to the index-matched liquid, and 1 mM of KCl was subsequently added to make the liquid conductive. The maximum emission wavelength for the dye DFSB-K175 corresponding to an excitation wavelength of 488 nm is 540 nm, while the maximum excitation and emission wavelengths for Alexa Fluor 647 are 653 nm and 669 nm respectively. For confocal imaging we use a Nikon A1 inverted line scanning confocal microscope with excitation lasers at 488 nm (for DFSB-K175 in the substrate) and 638 nm (for Alexa Fluor 647 in the drop), and with a 60x water immersion objective with numerical aperture (NA)=1.2. The emissions from the two dyes are collected using band filters in the range 500 nm to 550 nm for DFSB-K175, and in the range of 663 nm to 700 nm for Alexa Fluor 647. 3D confocal scans are performed in the immediate vicinity of the contact line such that the surface of the gel wetted by the droplet as well as the dry side are simultaneously visible. The 3D confocal scans are post-processed using ImageJ, and the xz slices are analyzed using an in-house MATLAB code to evaluate the deformation characteristics of the soft gel.

\section{Macroscopic contact angles}

\begin{figure}[h]
\centering
\includegraphics[width=86mm]{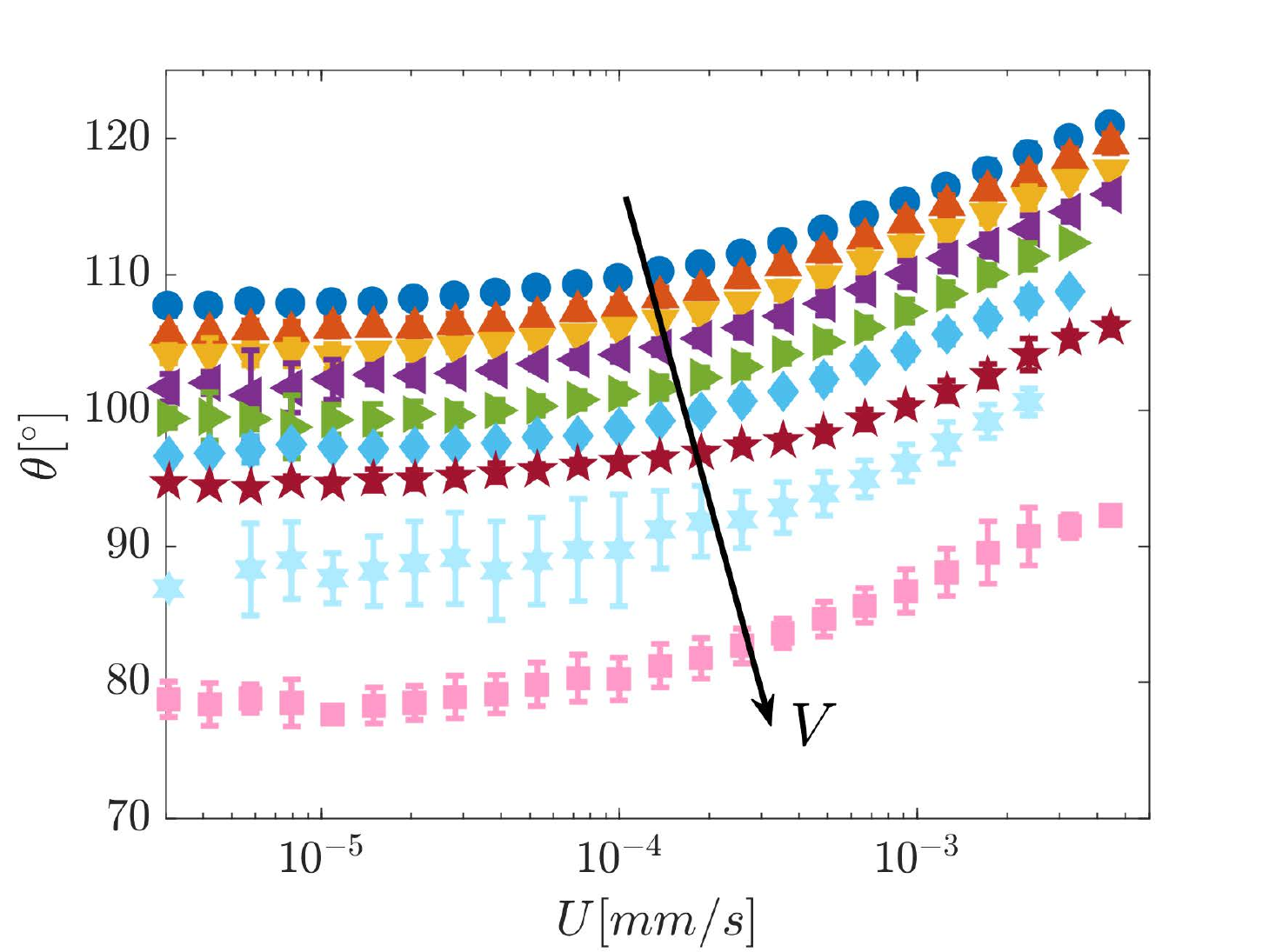}
\caption{Liquid contact angle $\theta$ as a function of contact line speed $U$ for increasing applied voltage. The data are obtained at $0$V (blue), $50$V(red), $75$V (yellow), $100$V (purple), $125$V (green), $150$V (light blue), $175$V (maroon), $200$V (sky blue), $250$V (pink).}
\label{fgr:theta_speed}
\end{figure}

The measurements of the static and dynamic contact angles are summarised in figure \ref{fgr:theta_speed}. We report the macroscopic liquid angle $\theta$ as a function of the instantaneous contact line speed $U=dr/dt$. The various datasets correspond to different applied voltages $V$, increasing from $0$V (blue) up to $250$V (pink). The plot shows that the contact angle $\theta$ increases as the speed $U$ increases, as is expected for advancing contact angles. For any given velocity, we clearly observe that $\theta$ decreases upon increasing the voltage $V$. The equilibrium angle at zero velocity, which we denote by the Lippmann angle $\theta_L$, is seen to decrease with $V$. The datasets at varying voltage all lie parallel to one another, as will be explained in more detail below. 

\subsection{Statics}

\begin{figure}
\centering
\includegraphics[width=86mm]{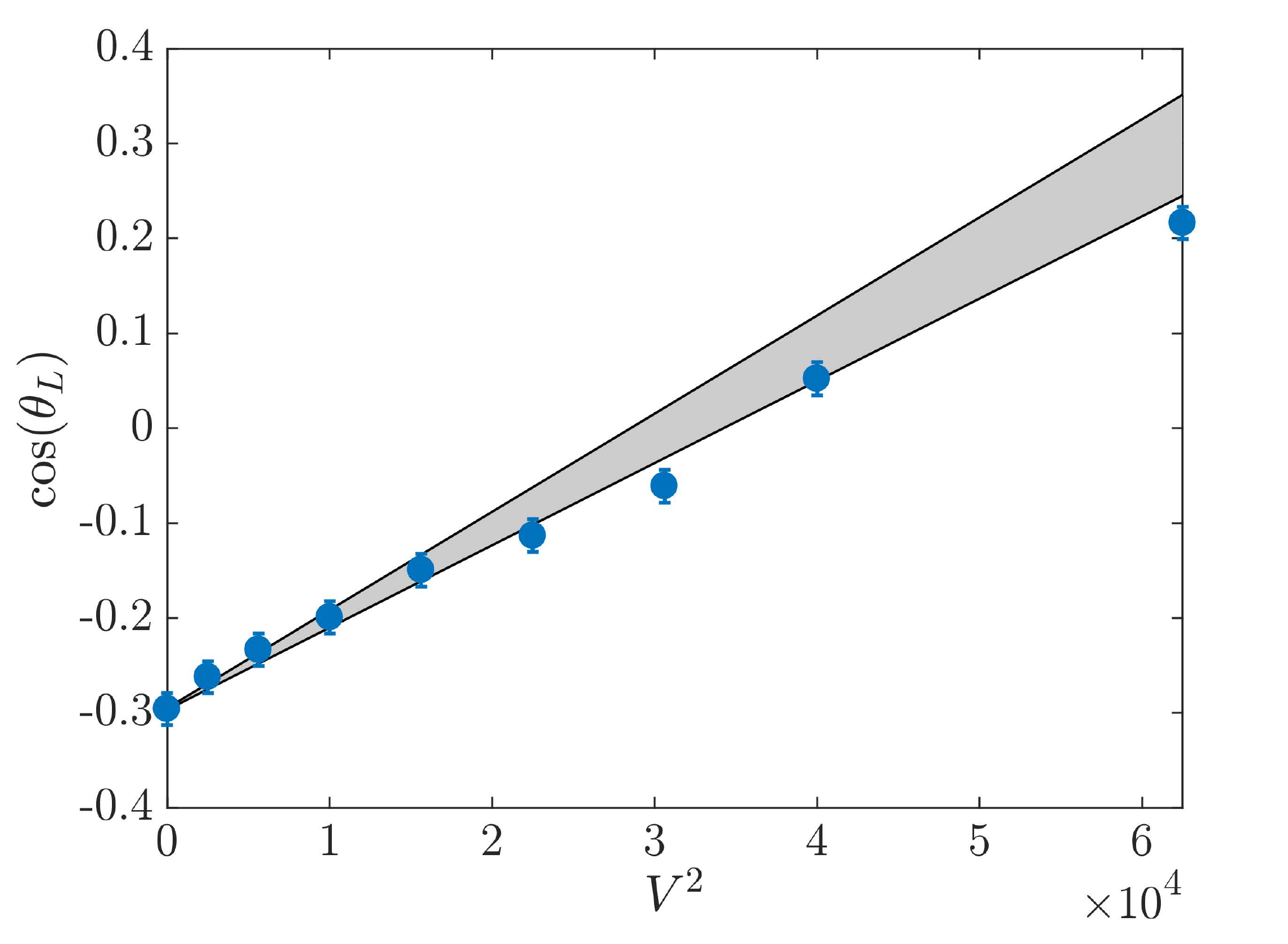}
\caption{Equilibrium angles as a function of the applied voltage squared. The shaded area shows the result of equation (\ref{eq:electrowetting}) based on the independently measured values of $\epsilon$, $h$ and $\gamma$.}
\label{fgr:EW_curve}
\end{figure}

We first analyse the measured value of the static angle $\theta_L$ as a function of voltage $V$. Given the Young-Lippmann equation (\ref{eq:electrowetting}), figure \ref{fgr:EW_curve} reports $\cos \theta_L$ as a function of $V^2$. The experimental data are shown as markers, and clearly exhibit a linear relation. Interestingly, the agreement of the experimental data with the Young-Lippmann equation is fully quantitative, even though the latter does not take into account elastic deformations. Using the calibrated values for $d$ and $\epsilon$, the slope predicted by (\ref{eq:electrowetting}) gives $9.5\pm0.9\times10^{-6}V^{-2}$, which is denoted by the shaded grey area in the figure. Hence, we conclude that the statics of electrowetting, viewed from the macroscopic perspective, is not affected by the substrate's deformability. Only at high voltages, the data lie slightly below the Young-Lippmann prediction, which could be the onset of the phenomenon of contact angle saturation \cite{Mugele2005,mugele2018electrowetting}.

To explain why the Young-Lippmann equation (\ref{eq:electrowetting}) still holds for soft dielectrics we revisit the macroscopic argument and estimate how this is altered by elasticity. The idea behind the macroscopic reasoning is that one can ignore the intricate structure of the three phase contact line (sketched e.g. in figure~\ref{fgr:schematic}c), by applying a global displacement $dx$ of the solution. Since the energy stored near the contact line remains unaffected by a global displacement, one only needs to consider the change in energies (capillary, electrostatic and elastic) at large distances from the contact line. One effectively exchanges a bit of dry surface, of area $dx$ times the length of the contact line, by a wetted region of the same area. On the dry part there is no Maxwell stress outside the contact line region, and hence there is no elastic deformation. Below the drop, however, there is a homogenous Maxwell stress $p_M = \frac{\epsilon \epsilon_0 V^2}{2d^2}$. On a compressible layer of Poisson ratio $\nu$, assuming linear elasticity, this leads to a compression $\Delta d$ of the layer

\begin{equation}
\frac{\Delta d}{d} = \left( \frac{1-2\nu}{1-\nu} \right) \frac{p_M}{2G} 
= \left( \frac{1-2\nu}{1-\nu} \right)  \frac{\epsilon \epsilon_0 V^2}{4d^2G} .
\end{equation}
The associated elastic energy per unit surface then can be written as $\frac{1}{2}p_M \Delta d$, leading to an increase of the effective surface energy per unit wetted area. Hence, the modified Young-Lippmann equation becomes

\begin{eqnarray}
\label{eq:electrowettingbis}
\cos \theta &=&\frac{\gamma_{SV}-  \left( \gamma_{SL}-\frac{\epsilon\epsilon_0 V^2}{2d} 
+ \frac{p_M \Delta d}{2} \right)}{\gamma} \nonumber \\
&=& \cos \theta_{L}  -  \frac{1}{4}\left( \frac{1-2\nu}{1-\nu} \right) \frac{p_M^2d}{G\gamma}.
\end{eqnarray}
In general, there is thus an elastic correction to $\theta_L$, which scales as $p_M^2 \sim V^4$. Importantly, however, our gel is nearly perfectly incompressible, with $\nu\approx 1/2$. Owing to the prefactor $1-2\nu$ in (\ref{eq:electrowettingbis}), this implies that there is no departure from the Yound-Lippmann angle $\theta_L$. The physical reason for the absence of an elastic correction is that the \emph{homogeneous} Maxwell stress below the drop (far away from the contact line) does not induce any deformation of the incompressible layer. By consequence no elastic energy is stored far away from the contact line, and thereby the macroscopic contact angle is unaffected by incompressible elasticity.

\subsection{Dynamics}\label{sec:dynamics}

\begin{figure}
\centering
\includegraphics[width=86mm]{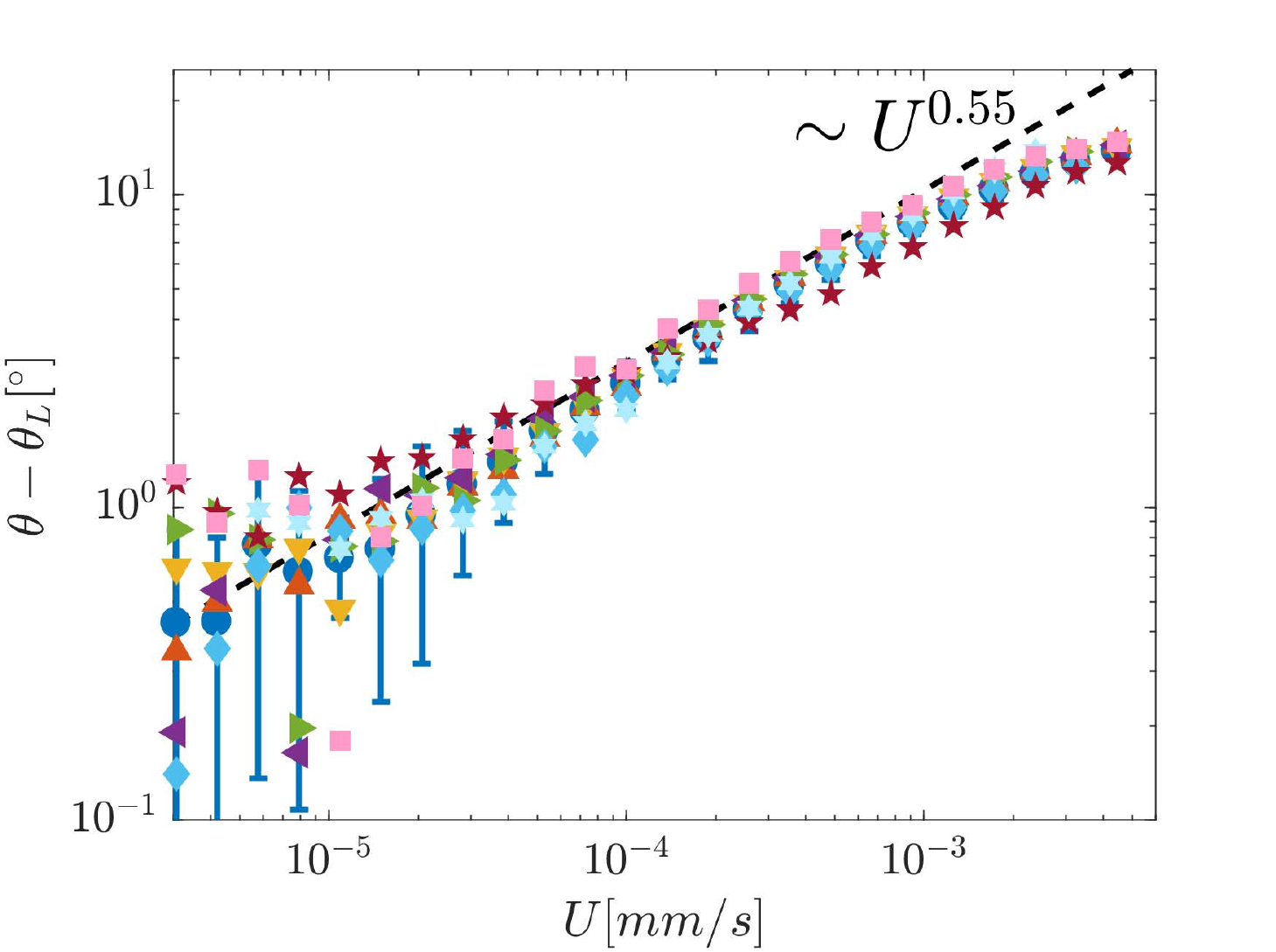}
\caption{Change of liquid contact angle from the equilibrium angle, $\theta-\theta_L$, as a function of contact line speed $U$. The data for different applied voltages collapse. Only the errorbars for the $0$V are shown, and the same color coding as figure \ref{fgr:theta_speed} is used.}
\label{fgr:dtheta_speed}
\end{figure}

Now we turn to the question on how electrostatic effects alter the contact line dynamics. For this we consider $\theta-\theta_L$,  the change of contact angle with respect to the static $\theta_L$,  as a function of contact line velocity $U$. The result is shown in figure \ref{fgr:dtheta_speed}, on a log-log plot. It can be clearly seen that all the curves collapse. This implies that the electrowetting effect is captured by the change in $\theta_L$, but that the subsequent dynamics is the same for all voltages. The experimental data exhibits a power law dependence, where the dashed line shows a fit $\sim U^{0.55}$. This behaviour was extensively studied in the case without electrowetting \cite{CGS96,LALLang96,KarpNcom15,zhao2018geometrical}, which is recovered here at zero voltage (blue data). The power-law directly reflects the frequency-dependence of the loss modulus of the layer, $G'' \sim \omega^n$, which indeed has $n\approx 0.55$. 

The scaling $\theta-\theta_L \sim U^n$ can in fact be derived from a macroscopic energy balance, equating the viscoelastic dissipation inside the solid to the power injected by capillary forces \cite{LALLang96}. We can thus inquire how this balance is affected by the presence of a voltage, and rationalise the collapse observed in figure \ref{fgr:dtheta_speed}. The electro-capillary power can be written as $\gamma U ( \cos \theta_L - \cos \theta) \simeq \gamma U \sin \theta_L \, (\theta - \theta_L) $, where by using $\theta_L$ we implicitly account for electrostatics. In principle, there is a dependence on voltage through $\theta_L$, but in our experiments $\sin \theta_L$ has little variation, only by about 5\%. The viscoelastic dissipation due to the motion of the ridge can be computed once the shape of the ridge is known \cite{Karpitschka2018}. For a given shape this leads to a dissipated power $\sim U^{n+1}$, where the prefactor depends on the wetting ridge morphology. Changes in this morphology, either due to large velocity or due to the application of a voltage, then alter the pure scaling law. This effect can be seen in figure \ref{fgr:dtheta_speed}, where at high velocity $U$ the data approaches a saturation \cite{KarpNcom15} -- this is due to a change in shape of the wetting ridge at high $U$. In principle the wetting ridge shape will also be different when a voltage is applied. However, the data in figure \ref{fgr:dtheta_speed} suggest that this change is not sufficient to have a measurable effect on the spreading dynamics. This will indeed be confirmed below by confocal microscopy measurements. 

\section{The wetting ridge}

\begin{figure}
\centering
\includegraphics[width=86mm]{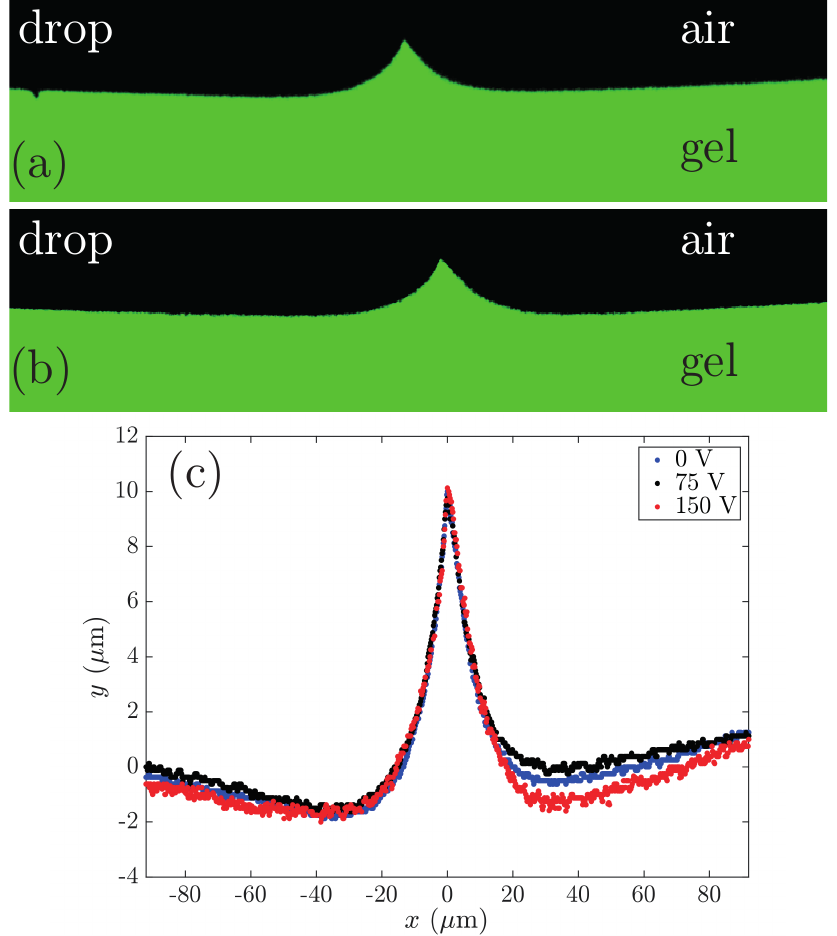}
\caption{(a,b) Confocal microscopy images of the wetting ridge below the contact line of a liquid drop on gel, at $0$ V (panel a) and $150$ V (panel b). (c) Profiles of the wetting ridge extracted from the confocal measurements at $0$V (blue), $75$ V (black), and $150$V (red).}
\label{fgr:confocal}
\end{figure}

Since both the elastocapillarity and electrostatic effects act in close vicinity of the contact line, we zoom in on the details at the contact line using confocal microscopy. Our results of the confocal measurements of the wetting ridge are reported in figure \ref{fgr:confocal}. We show two wetting ridge shapes at equilibrium: Panel (a) corresponds to a case without any electrowetting, while panel (b) corresponds to $150$V. Both images show a confocal slice. The extracted profiles of the wetting ridge for different applied voltages are shown in figure \ref{fgr:confocal}c, for  0 V (blue), 75 V (black) and 150 V (red).

We observe that both with and without an applied voltage the gel has a dimple to the left and to the right of the wetting ridge. This is a consequence of the incompressibility of the gel -- the volume inside the wetting ridge is directly drawn from the adjacent region. The minimum of the dimple is at a distance comparable to the thickness of the gel~\cite{karpitschka2016liquid}. 
 
When comparing the shapes of the wetting ridge without and with electrowetting, one observes hardly any difference when a voltage is applied (at the wetting ridge, the differently coloured markers overlap in figure \ref{fgr:confocal}c). This shows that on the scale of the wetting ridge the capillary forces, which do induce substantial elastic deformation, are more prominent than electrostatic forces. Indeed, the singularity of the Maxwell stress near the contact line is substantially weaker than for the classical case of a droplet on a rigid dielectric. This is because the angle between the liquid-solid and liquid-vapor interface is much larger owing to the elastic deformation, thereby reducing the degree of the electrostatic singularity. 
Furthermore, the tip of the wetting ridge shows no measurable change in the solid opening angle. The constant tip angle suggests that the Neumann law is still applicable at the ridge tip, representing a balance of the three involved surface tensions. This nicely illustrates that at the ultimate scale, neither Maxwell stress or bulk elasticity can compete with surface forces -- a result that is known for both electrowetting~\cite{Buehrle2003} and for soft wetting~\cite{style2017elastocapillarity, Snoeijer2018}. 

Finally, it must be also noted in figure \ref{fgr:confocal}c that there is no significant change in the height of the wetting ridge with the applied electrical voltage. This is because the vertical forcing by the drop $\gamma \sin \theta$ changes merely by $\sim 4\%$ with the applied voltage (over a range of 150 V), and hence, fails to create any significant change in the wetting ridge height. In essence, the wetting ridge shape does not change significantly with electrowetting, which further substantiates our interpretation of the spreading dynamics as discussed in Section \ref{sec:dynamics}.

\section{Discussion}

In summary, we have investigated the macroscopic contact angles for electrowetting on a soft dielectric layer, both in static and in dynamic conditions. It is found that the statics of electrowetting is not affected by elasticity -- the macroscopic contact angle is still given by the Young-Lippmann equation. We have attributed this absence of elastic effects on $\theta_L$ to the incompressibility of the dielectric layer, preventing the storage of elastic energy far away from the contact line. However, the viscoelasticity of the dielectric has a dramatic effect on the electrowetting dynamics. The dynamic contact angle, quantified by the change of the angle with respect to $\theta_L$, is completely dictated by the viscoelastic dissipation inside the solid, rendering it much slower than normal electrowetting dynamics. Phrased differently, one could also state that the electrostatics does not influence the dynamics of spreading on soft surfaces, apart from changing the equilibrium angle. 

Subsequently, we have qualitatively explored the structure of the wetting ridge, comparing shapes with and without an applied voltage. We have shown that there is no significant change in the wetting ridge shape with the applied voltage, which explains the collapse of the dynamical spreading experiments in Sec.~\ref{sec:dynamics}. It would be of interest to provide a more detailed study of mechanics on the scale of the wetting ridge, in a regime where electrostatic and elastic forces are both prominent. This could be done in the form of a more extensive experimental study using confocal microscopy, in combination with explicit mechanical calculations that predict the shape of the wetting ridge in presence of external electrical stresses. 
Furthermore, it will be worthwhile to investigate whether the independence of EW and the electrostatics induced deformation of the wetting ridge can be extended to liquid-infused surfaces \cite{hao2014electrowetting}, where the spreading dynamics is generally often dictated by the dissipation in the liquid wetting ridge \cite{keiser2017drop}. However, the understanding of soft electrowetting presented here is sufficient to design new applications involving electrowetting induced droplet manipulation on soft dielectrics instead of rigid ones as routinely done so far.

\emph{Acknowledgments.~---~} 
We gratefully acknowledge discussions with B. Andreotti and S. Karpitschka. 
MvG and JHS are financially supported by the ERC (the European Research Council) Consolidator Grant No. 616918.

\footnotesize{
\bibliography{electro_elasto} 
\bibliographystyle{rsc} 
}

\end{document}